\documentclass[aps,twocolumn,nofootinbib]{revtex4}

\usepackage{graphicx} 
\usepackage{epstopdf} 
\usepackage{url}
\usepackage{float}
\usepackage{fancyhdr}
\usepackage{hyperref}
\usepackage{wrapfig}

\fancypagestyle{firstPage}
{
	\pagestyle{fancy}
	\lhead{Amanda Steinhebel}
	\chead{}
}

\begin{document}

\title{Studies of the Response of the SiD Silicon-Tungsten ECal}

\author{%
\textsc{Amanda Steinhebel and James Brau}\\
\normalsize \textit {University of Oregon, Center for High Energy Physics, 1274 University of Oregon Eugene, Oregon 97403-1274 USA} \\
}
\date{\today}

\begin{abstract}
\begin{center}
{\small \textbf{Talks presented at the International Workshop on Future Linear Colliders (LCWS16),  \\ Morioka, Japan, 5-9 December 2016. C16-12-05.4}}
\end{center}
~\newline
\indent Studies of the response of the SiD silicon-tungsten electromagnetic calorimeter (ECal) are presented. Layers of highly granular (13 mm$^2$ pixels) silicon detectors embedded in thin gaps ($\sim$ 1 mm) between tungsten alloy plates give the SiD ECal the ability to separate electromagnetic showers in a crowded environment. A nine-layer prototype has been built and tested in a 12.1 GeV electron beam at the SLAC National Accelerator Laboratory. This data was simulated with a Geant4 model. Particular attention was given to the separation of nearby incident electrons, which demonstrated a high (98.5\%) separation efficiency for two electrons at least 1 cm from each other. The beam test study will be compared to a full SiD detector simulation with a realistic geometry, where the ECal calibration constants must first be established. This work is continuing, as the geometry requires that the calibration constants depend upon energy, angle, and absorber depth. The derivation of these constants is being developed from first principles.
\end{abstract}

\maketitle 


\section{Introduction}
SiD is one of two detectors under consideration for use in the International Linear Collider (ILC) \cite{tdr4}. Its electromagnetic calorimeter (ECal) is a sampling calorimeter constructed of alternating layers of silicon diodes and DENS-24, a tungsten alloy used as an absorber. The silicon diode pixels individually record charge deposited by particles from the electron-positron collision. Each pixel is individually read out by a KPiX chip \cite{kpix}.

 This document summarizes work done at the University of Oregon regarding the response of the KPiX readout chip and geometry of the SiD ECal. 

\section{KPiX Background}
\begin{figure}[t!]
	\begin{center}
		\includegraphics[width=0.6\linewidth]{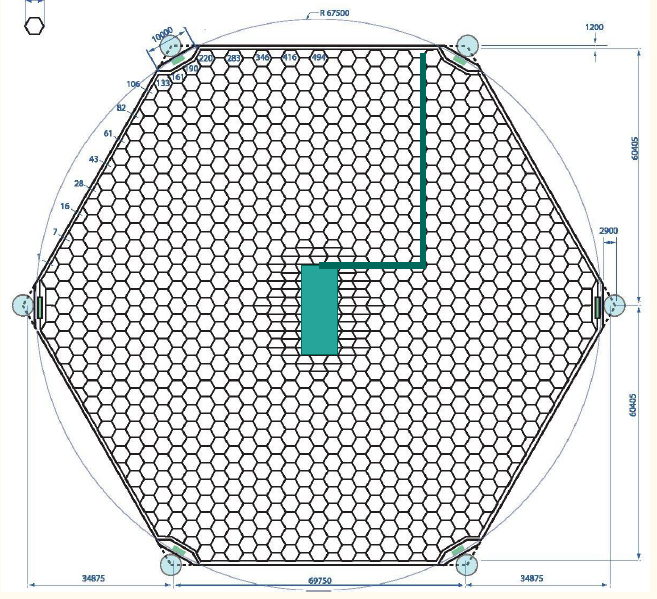}
		\caption{The engineering schematic of one ECal silicon wafer, with the position of the KPiX readout chip shown in green \cite{tdr4}.}
		\label{wafer}
	\end{center}
\end{figure}

Thirty-one 0.3 mm thick silicon layers are created from a tiling of hexagonal wafers each containing 1024 individual 13 mm$^2$ pixels. As a photon or electron from the collision passes through the ECal,  the tungsten layers produce showering. The silicon layers then measure any charge deposited on them as the shower progresses through the calorimeter. One KPiX readout chip is bump-bonded to the center of each silicon wafer (Fig.~\ref{wafer}) that is connected via channels to each pixel on the chip. In this way, measurements from every individual pixel are read out, providing measurements for every 13 mm$^2$. The use of KPiX allows for thin sampling layers of 1.25~mm.

\begin{figure}[t!]
	\begin{center}
		\includegraphics[width=0.7\linewidth]{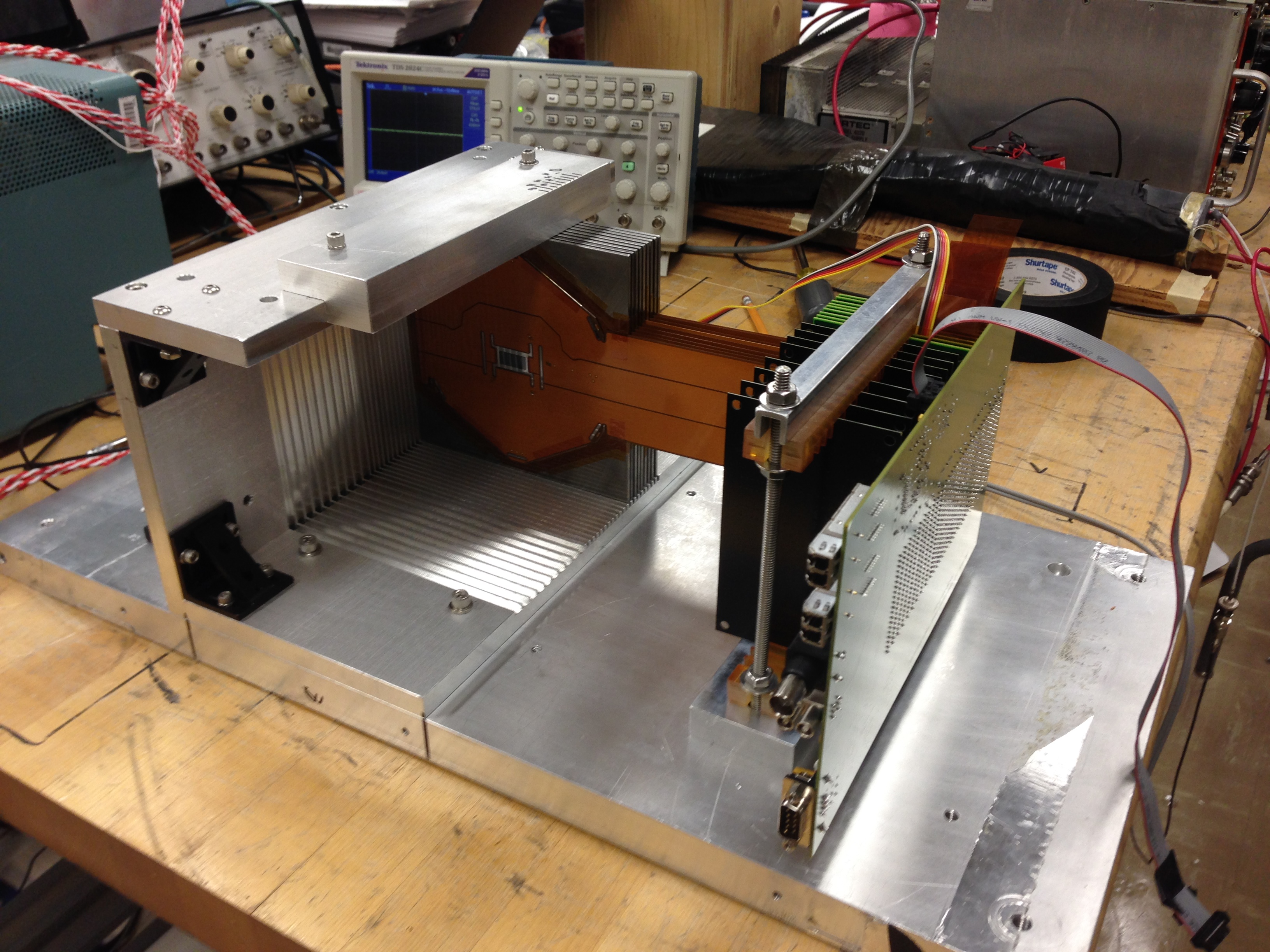}
		\caption{The ECal prototype setup at SLAC, in a silicon-first arrangement. [Photo credit: Marco Oriunno]}
		\label{setup}
	\end{center}
\end{figure}

Prototype versions of silicon wafers mounted with KPiX chips were tested at SLAC National Accelerator Laboratory in 2013 to test the response of the calorimeter \cite{lcws13}. The prototype calorimeter consisted of nine repeated alternating layers of silicon wafers and 2.5 mm DENS-24 plates (Fig.~\ref{setup}), for a total expanse of 5.8 radiation lengths. A 12.1 GeV electron beam was directed through the prototype calorimeter, and the KPiX response was recorded. The alternating silicon/tungsten pattern allowed for testing both silicon- and tungsten-first setups.
\begin{figure}[t!]
	\begin{center}
		\includegraphics[width=0.9\linewidth]{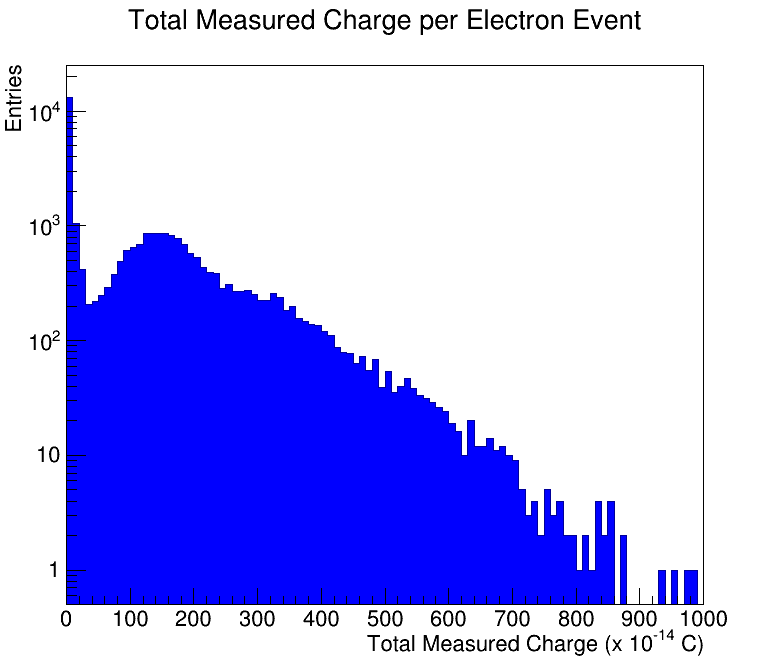}
		\caption{Total measured charge per event from the beam test where the silicon layer was placed first. Note the large peak of low energy events, as well as clear peaks around intervals of 150$\times 10^{-14}$ C indicating electron events.}
		\label{testBeamOrig}
	\end{center}
\end{figure}
  
\section{Test Beam Modeling Studies}
Before the analysis, the beam test data was cleaned. This cleaning included the removal of ``monster events", or events in which all pixels unrealistically reported a large amount of deposited charge. This phenomenon has since been understood. After the monster events were removed, a large number of low energy events remained from the data set of more than 30,000 events (Fig.~\ref{testBeamOrig}). These are accompanied by peaks at intervals of 150$\times 10^{-14}$ C, indicating electron events. The peaks at higher measured charge imply multiple electron events. 

Many low energy events are soft photon contamination from the electron beam. Setting a higher threshold on the recorded charge would eliminate the consideration of this contamination, but would also neglect low shower-energy electron events. In order to clean out only the contamination, an algorithm was designed to categorize showers. In this way, ``photons" could be separated from ``electron" showers and eliminated. 

A simple categorization technique is to count how many layers of silicon record hits in a given event. Roughly $45\%$ of events only contains hits in one layer. This type of event is characteristic of soft photon contamination, and can be immediately removed from consideration.

A weighting algorithm was developed to further categorize showers. The silicon layers were labeled from $1\rightarrow 9$, with layer one being the first silicon layer the beam encounters. Then, the ratio
\begin{equation}
	R=\frac{\sum_hL_h^2\mathcal{C}_h}{\sum_h\mathcal{C}_h}
	\label{weight}
\end{equation}
was calculated, where $\mathcal{C}_h$ is the measured charge for a given hit and $L_h$ is the layer number of the hit, summed over all hits $h$. If for some layer there were no hits, a weight of $5\times 10^{-14}$ C was inserted. This is roughly the charge that a minimum ionizing particle would deposit. 
\begin{figure}[b!]
	\begin{center}
		\includegraphics[width=0.9\linewidth]{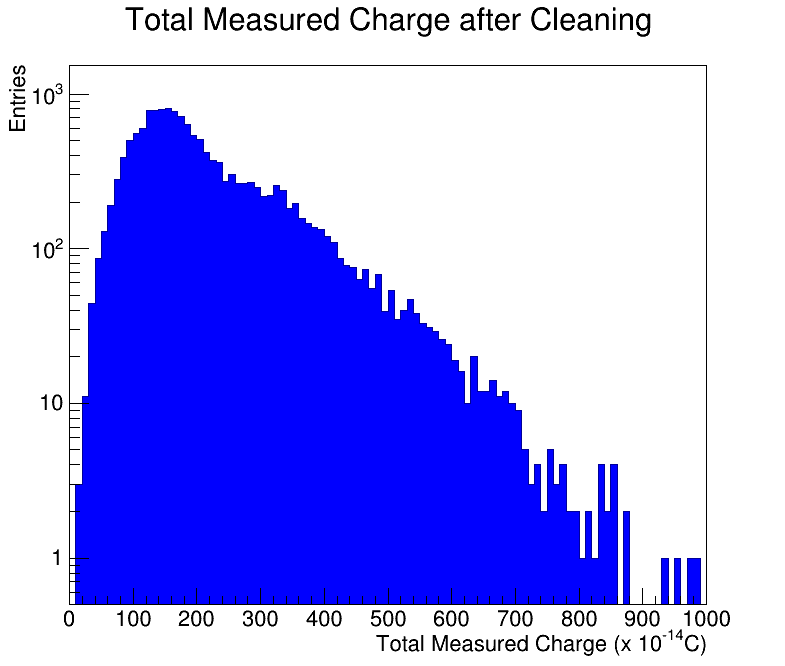}
		\caption{Deposited charge data from the beam test where the silicon layer was placed first, after removing soft photon contamination.}
		\label{cleaned}
	\end{center}
\end{figure}
\begin{figure}[b!]
	\begin{center}
		\includegraphics[width=0.9\linewidth]{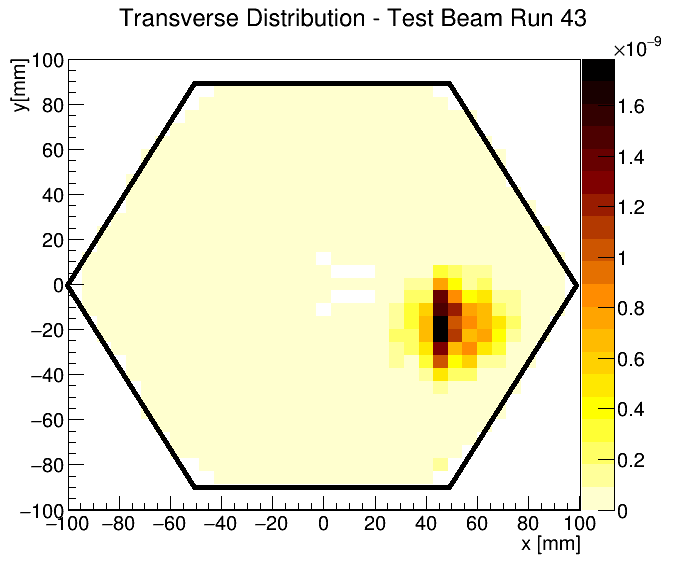}
		\caption{Profile of the position of hits along the ECal prototype. The intensity of color indicates the amount of measured charged summed over all nine layers at that $xy$ position, with a solid line indicating the edge of the wafer. Incident electrons simulated with Geant4 mimic this distribution.}
		\label{tbPosition}
	\end{center}
\end{figure}

Photon events tend to appear early and only in a few layers of the detector, causing  $R$ to be small due to the quadratic dependence on layer number. Similarly, electron showers peak later in the detector and drive $R$ up. A cut on $R$ was then applied, and events with $R$ less than the cut value were disregarded as photon contamination.

After this procedure, nearly $50\%$ of all events from the data set were removed. The resulting data set (Fig.~\ref{cleaned}) retains low shower-energy electron events while the large photon contamination peak has clearly been removed.

A Geant4 simulation was created to model the beam test scenario. The simulation consisted of 8,000 single electron events transversely distributed in the calorimeter to match the beam test data (Fig.~\ref{tbPosition}).

\begin{figure}[b!]
	\begin{center}
		\includegraphics[width=0.9\linewidth]{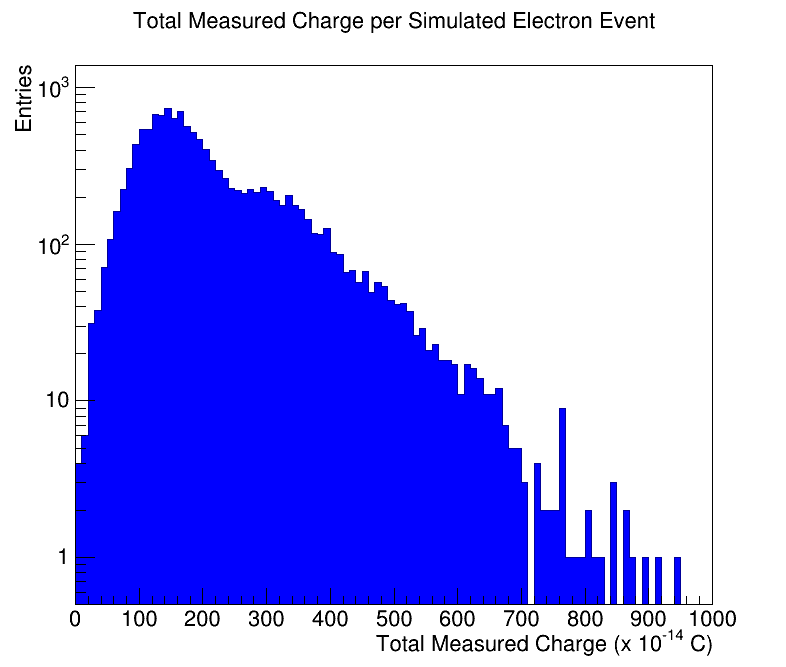}
		\caption{Geant4 simulated data designed to match the beam test prototype.}
		\label{simOrig}
	\end{center}
\end{figure}

The collection of single-electron events was then used to create a Poisson distribution of multi-electron events by overlaying multiple single-electron events. In order to simulate inactive pixels observed during the beam test, $10\%$ of the pixels of each layer were randomly removed. The resulting data set is shown in Fig.~\ref{simOrig}.
\begin{figure}[b!]
	\begin{center}
		\includegraphics[width=0.9\linewidth]{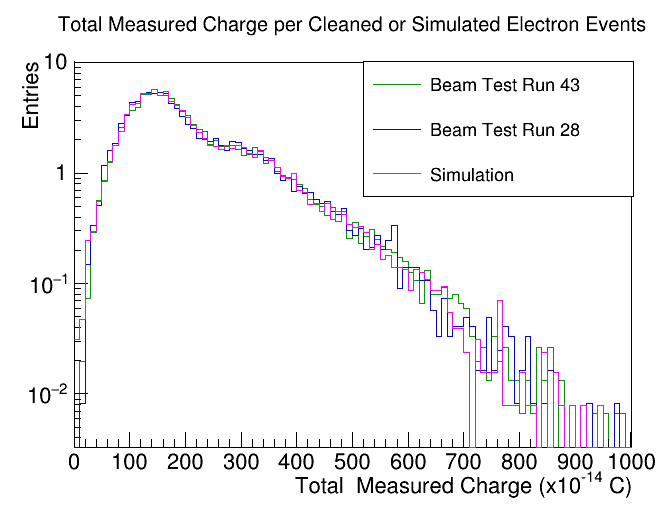}
		\caption{Two silicon-first prototype runs (labled ``Run 28" and ``Run 43") match very well with Geant4 simulated data when the total measured charge in each event is compared after the prototype data sets are cleaned of soft-photon contamination.}
		\label{combo}
	\end{center}
\end{figure}

\begin{figure*}
	\begin{center}
		\includegraphics[width=0.9\linewidth]{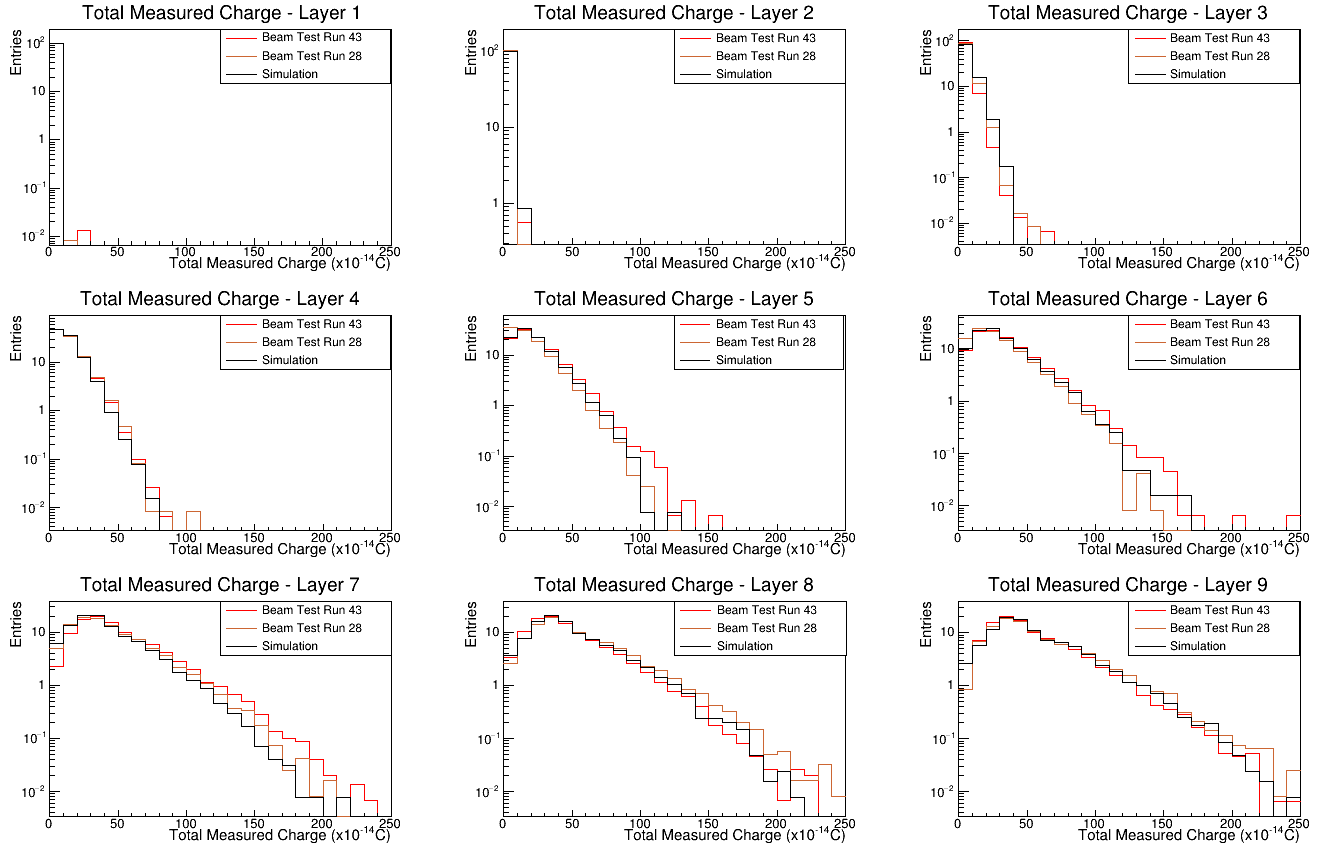}
		\caption{Two silicon-first prototype runs (labeled ``Run 28" and ``Run 43") match very well with Geant4 simulated data when the total measured charge in each layer of each event is compared after the prototype data sets are cleaned of soft-photon contamination.}
		\label{layers}
	\end{center}
\end{figure*}

Once all runs were normalized to one hundred events, the simulated and collected data agree well (Fig.~\ref{combo}). This agreement holds not only for the total measured charge in each event, but for the total measured charge in each layer of the prototype detector of each event as well (Fig.~\ref{layers}).

\section{Shower Separation Efficiency Studies}
The high spatial granularity provided by the silicon pixels and KPiX can help distinguish the showers of two nearby particles. Since two-electron events were clearly observed in the beam test prototype run, a study of the prototype's ability to separate showers was done. An algorithm was created to count the number of incident particles detected in each event. This algorithm is simple in nature, but robust enough to examine data from both the beam test prototype and Geant4 simulation. It requires inputs that describe the geometry of the silicon wafer, including which pixels border which other pixels, and the charge measured in each pixel. The algorithm simply examines each layer of each event and determines local maxima of charge deposits. It then compares the position of this maximum against all other layers, and requires that the same pixel location be a local maximum in at least four layers. If this condition is met, then an electron event is counted. In this way, the algorithm can account for multiple electrons occurring within one event but it also biases against late forming showers that do not develop fully enough to create four layers with notable maxima. 

Occasionally, a shower maximum occurred near the border of two pixels. In this case, the recorded local maximum was equally likely to be located in either pixel. This fooled the algorithm into tagging two incident particles though there was only truly one. The counting of more particles than the true number of incident particles is considered``over-counting". The algorithm corrects for this by disallowing the presence of maxima in neighboring pixels.

\begin{figure}
	\begin{center}
		\includegraphics[width=\linewidth]{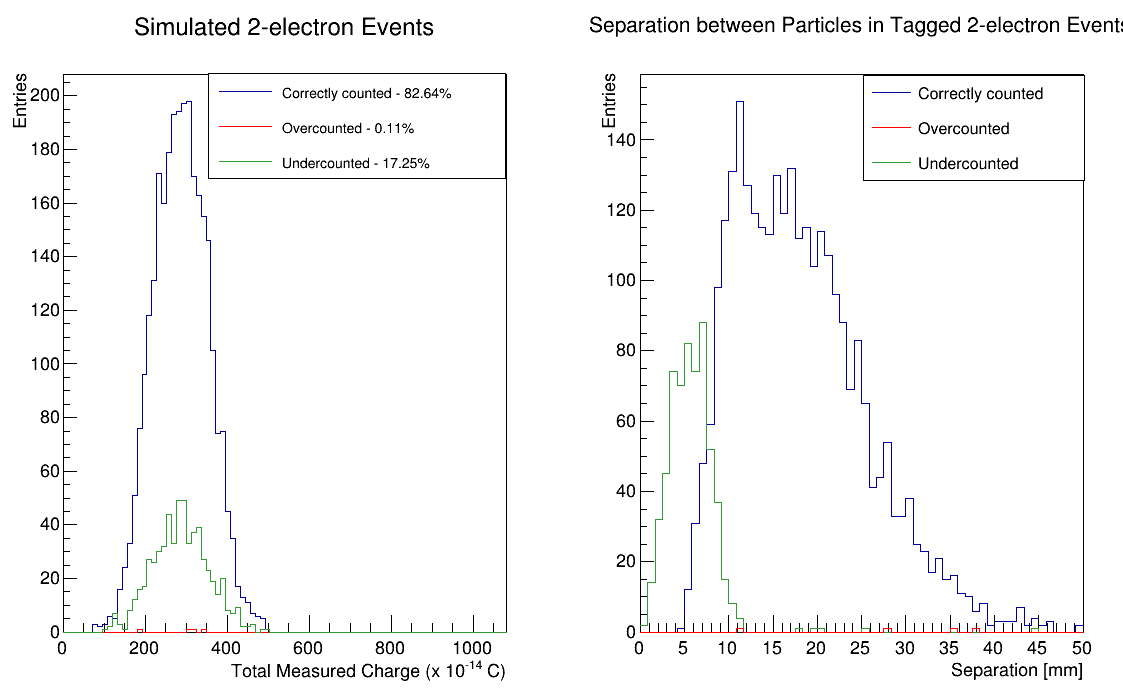}
		\caption{Performance of the electron counting algorithm when measured with simulated two-electron events. The algorithm tended to correctly count two electrons (blue lines), but when miscounting under-counting (green lines) was more common than over-counting (red lines). Under-counted events tended to occur when the two electrons were spatially close (less than 1 cm apart).}
		\label{2count}
	\end{center}
\end{figure}

\begin{figure}[h!]
	\begin{center}
		\includegraphics[width=0.9\linewidth]{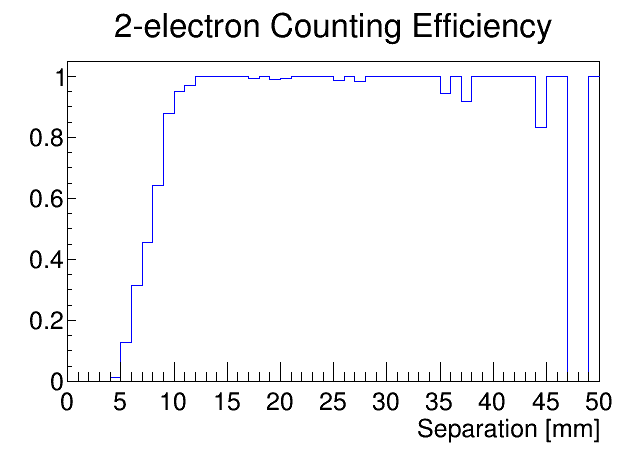}
		\caption{Efficiency of algorithm with simulated two electron events. \newline}
		\label{efficiency}
	\end{center}
\end{figure}

With simulated data, truth information regarding the number of incident electrons is available and the accuracy of the algorithm can be examined. Among simulated two-electron events, the algorithm correctly counted 82.6\% of events. 17.3\% of two-electron simulated events were under-counted (the algorithm detected less incident particles than there really were), meaning that if the algorithm miscounted it was far more likely to under-count than over-count (Fig.~\ref{2count}).
When events were incorrectly under-counted, the two electrons tended to be less than 1 cm apart. The algorithm counted two-electron events with an average efficiency of $98.5\%$ when the incident electrons were separated by more than 1 cm (Fig.~\ref{efficiency}).

\begin{figure}
	\begin{center}
		\includegraphics[width=0.9\linewidth]{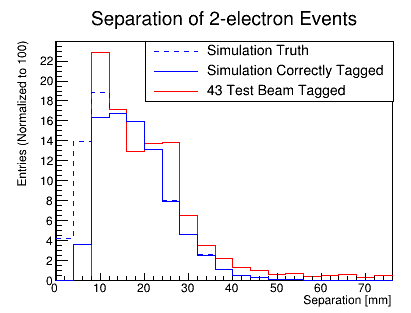}
		\caption{The algorithm tags two-electron events from the beam test and simulated data at similar rates. The tagging is less accurate when the electrons are separated less than 1~cm. }
		\label{normSep}
	\end{center}
\end{figure}

The algorithm can also analyze data from the beam test prototype, though truth information is unknown. Events the algorithm tags from the beam test data as ``two-electron events" compare appropriately to those
\newline
\newline
\newline
\newline
that the algorithm tags from the simulated data set (which can also be compared to simulation truth data) (Fig.~\ref{normSep}). This implies both that the simulation is correctly modeling the system, and that the algorithm can be trusted to identify multi-electron events in prototype data with nearly perfect efficiency provided that incident electrons are separated by more than 1~cm.

The ability to discern multiple electrons incident in similar spatial regions is important for reconstruction of particles using the ECal information - especially the ability to reconstruct boosted $\pi^0$ mesons from their decay products of two photons.

\section{ECal Overview}
\begin{figure}
	\begin{center}
		\includegraphics[width=0.9\linewidth]{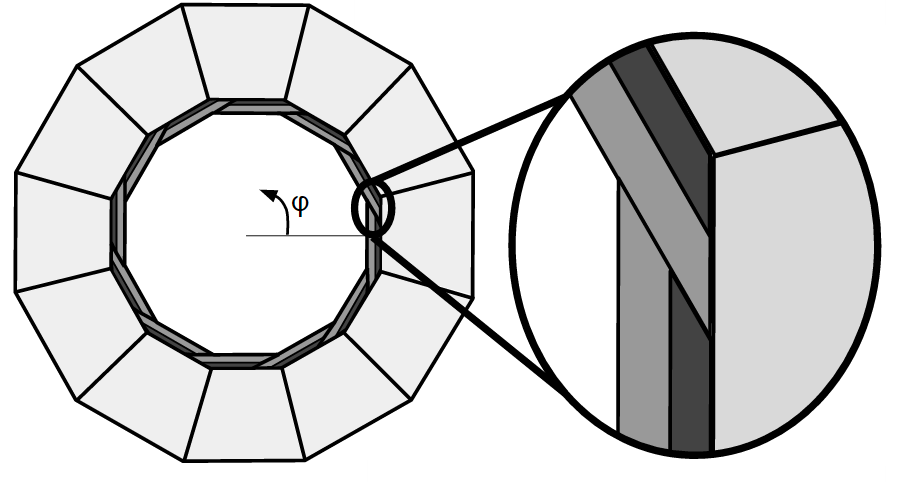}
		\caption{The SiD ECal (darker colors) is surrounded by the hadron calorimeter (lighter colored) and made of twelve overlapping trapezoidal modules to avoid projective cracks. The cutout image illustrates the overlap region of these ECal modules, where the darkest shade indicates areas with thick tungsten layers and the medium-colored shade indicates areas with thin tunsten layers. The full image, showing the view from the $xy$ plane with the $z$ dimension coming out of the page, also indicates the angle $\varphi$.}
		\label{geo}
	\end{center}
\end{figure}

In the SiD design, the ECal barrel sits between the vertex tracker and hadron calorimeter, at an inner radius of 1264~mm from the collision point with a $z$ extent of 3.53~m. It is made of twelve trapezoidal modules that extend the full $z$ length of the detector with overlapping ends to avoid projective cracks through the detector, creating a structure that is periodic in increments of $\pi/6$ radians  (Fig.~\ref{geo}  shows the view from the $xy$ plane with the $z$ dimension coming out of the page, also indicates the angle $\varphi$.). These trapezoids have a small inner angle of 30$^o$ \cite{tdr4}.
The region of overlap between two modules spans from $\varphi\in [(4.03+30n)^o,(15+30n)^o]$, where $n$ is an integer, $n=0,1,\dots,11$. This is approximately 30\% of the detector.

Each module consists of 31 layers of tiled silicon wafers and 30 layers of DENS-24 (as detailed in Section~II). The first layer of each module is a silicon tracking layer, followed by twenty iterations of 2.5~mm DENS-24 and a 1.25~mm gap in which the 0.3~mm silicon layer will reside. This is followed by ten iterations of 5 mm DENS-24 and the same 1.25~mm silicon-containing gap \cite{tdr4}. Therefore, the ECal begins and ends in a sensitive silicon layer. This structure is identically repeated for all twelve modules.

The combination of a nontrivial ECal geometry and unequal division of tungsten absorber throughout each module creates complicated calorimeter calibration and geometric effects. For example, a smaller subset of the overlap region, where  $\varphi\in [(8.786+30n)^o,(10.14+30n)^o]$, contains only thin layers of tungsten absorber, and no thicker 5 mm tungsten layers. This is the ``thin overlap region".

All the following studies were conducted using the full SiD simulation, $SiD\_o1\_v03$. Single photons are directed into the detector at various $\varphi$ angles incident to the ECal surface ($\theta=90^o$) and at initial energies of 10~GeV or 100~GeV. Only charge deposited in the silicon layers of the ECal barrel is considered.

\section{Geometry Studies}
\begin{figure}[t!]
	\begin{center}
		\includegraphics[width=0.9\linewidth]{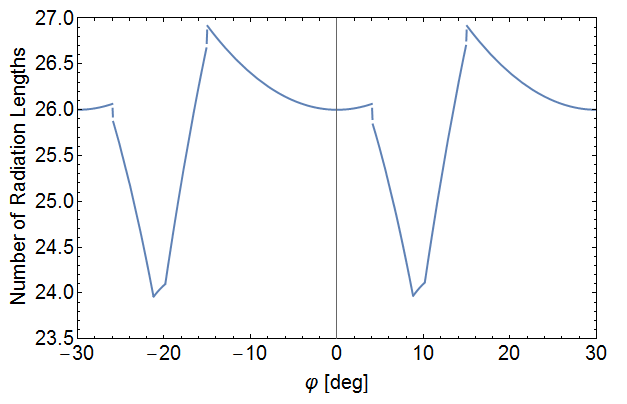}
		\caption{Total number of DENS-24 radiation lengths in the ECal barrel at different $\varphi$ positions. At normal incidence, there are 26 radiation lengths.}
		\label{radLenNo}
	\end{center}
\end{figure}

\begin{figure}[t!]
	\begin{center}
		\includegraphics[width=0.9\linewidth]{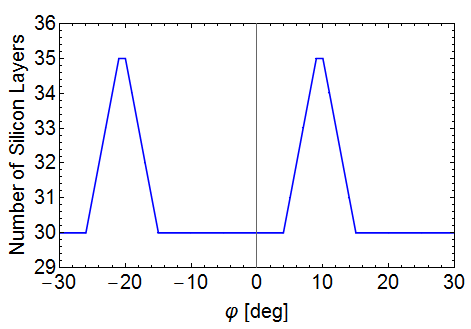}
		\caption{Total number of silicon layers in the ECal barrel at different $\varphi$ positions, excluding the initial silicon tracking layer. At normal incidence, there are 30 silicon layers.}
		\label{noLay}
	\end{center}
\end{figure}

The overlapping geometry of the ECal barrel creates various effects that must be taken into account in the calorimeter calibration, including a varied absorber depth and number of incident sensitive layers that depend on the angle $\varphi$ (Figs.~\ref{radLenNo} and \ref{noLay}). 
To investigate this, 500 events of 100~GeV and 10~GeV photons were run through the full SiD simulation at $\varphi~=~0^o,~3.25^o,~7.5^o,~9.3^o,~11.25^o,~15^o,~18.25^o,~26.25^o,$ and $30^o$. This surveys through one $30^o$ period of the detector, with two points in the overlap region ($\varphi~=~7.5^o~\mathrm{and}~11.25^o$) and one point in the thin overlap region ($\varphi=9.3^o$). 

The tracking layer of silicon at the beginning of each module is excluded from these studies. In overlap regions, this layer falls near the middle of the calorimeter and samples showers before they have traversed a full absorber layer. In this sense, the shower is being double-sampled due to the presence of this sensitive layer\footnote{This conclusion has led the SiD collaboration to consider designing the modules so that the tracking silicon layer is only present around the inner circumference of the ECal. Reducing the extent of this layer can be a cost-saving method.}.

\begin{figure}
	\begin{center}
		\includegraphics[width=0.9\linewidth]{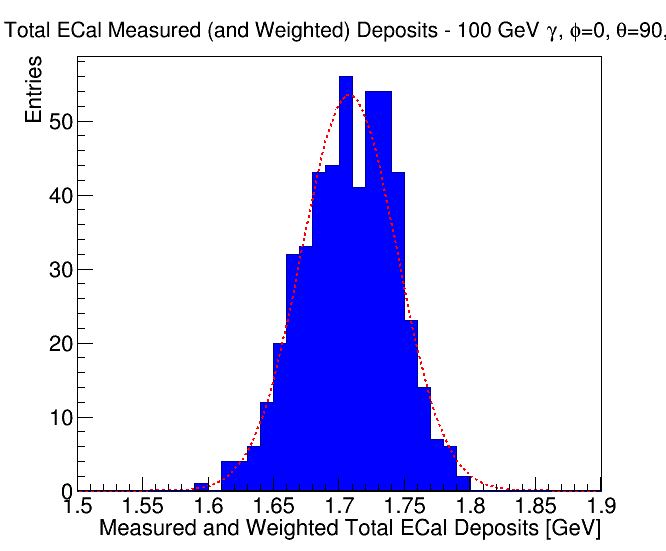}
		\caption{Total measured charge in the ECal, with deposits following thicker 5 mm absorber layers weighted by 2, for 500 photons with initial energy of 100 GeV.}
		\label{100GeV0phi}
	\end{center}
\end{figure}

At each $\varphi$ angle, a histogram was made of the total charge measured in the ECal barrel\footnote{Deposits in the ECal endcap were not considered in this study.}. Since the last ten silicon layers follow thicker tungsten layers than the first 20, deposits in these layers are weighted by a factor of two to account for the differing sampling fraction\footnote{The true value is 1.98, however using a factor of 2 agrees to within a few percent.}. Figure~\ref{100GeV0phi} shows an example of 500 events of 100 GeV photons with $\theta=90^o$ and $\varphi=0^o$. The data is fit with a Gaussian, giving an average standard deviation of 2.1\% of the mean for the 100 GeV events. 

\begin{figure}
	\begin{center}
		\includegraphics[width=\linewidth]{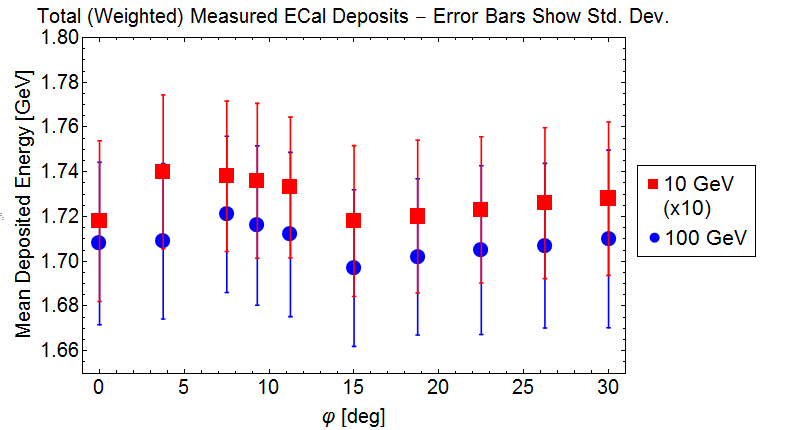}
		\caption{The average weighted measured charge of photon events of initial energies of 100 GeV and 10 GeV (scaled by ten) as a function of $\varphi$, or the angle around the SiD ECal. Error bars indicate the standard deviation that also supplied the plotted mean values.}
		\label{survey}
	\end{center}
\end{figure}

From analogous distributions for all $\varphi$ angles, the mean value from the Gaussian fit was recorded and plotted as a function of $\varphi$ with error bars representing the standard deviation. This was done for 100 GeV initial photon energies and 10 GeV initial photon energies (where the mean energies of the 10 GeV runs are scaled up by a factor of ten to compare to the 100 GeV runs) (Fig.~\ref{survey}).
\begin{figure}
	\begin{center}
		\includegraphics[width=\linewidth]{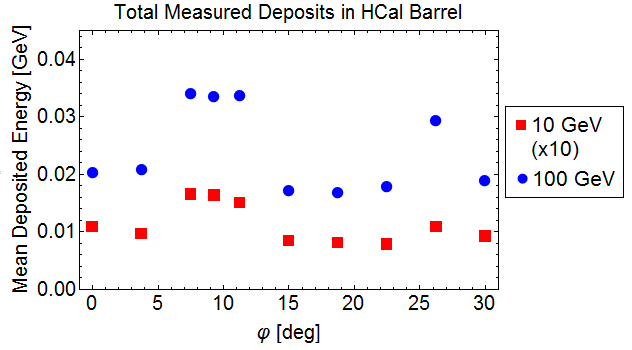}
		\caption{The average, uncalibrated measured charge in the hadron calorimeter of photon events of initial energies of 100~GeV and 10~GeV (scaled by ten) as a function of $\varphi$. The higher energy 100~GeV photons experience more leakage into the hadron calorimeter.}
		\label{hcal}
	\end{center}
\end{figure}

The standard distribution of measured charge remains fairly constant throughout the entire $\varphi$ range, with standard deviations of 100 GeV and 10 GeV runs in the range of 2.0\%~-~2.3\% and 5.7\%~-~6.6\% of the mean, respectively. This is seen even in the overlap region of $\varphi\in[4.03^o,15^o]$ where the showers experience higher sampling rates (see Fig.~\ref{noLay}). The 10 GeV runs, once scaled up by a factor of ten, have a slightly higher mean than the 100 GeV run due to lower leakage into the hadron calorimeter (Fig.~\ref{hcal}). Deposits in the hadron calorimeter notably see an increase in measured charge of more than $60\%$ in the overlap region of the ECal. This is due to the lower total radiation lengths within the this region (Fig.~\ref{radLenNo}), where the number of radiation lengths decreases from $26~X_0$ at normal incidence to a minimum of $23.7~X_0$ at $\varphi~=~(8.786n)^o$.

These effects are currently under investigation at the University of Oregon, as effort continues to optimize the SiD ECal design from first principles and to formulate calibration constants that include energy- and angular-dependence.

\section{Acknowledgments}
We would like to thank Jason Barkeloo, Teddy Hay, and Dylan Mead for their extensive previous work with the electron-counting algorithm and SLAC beam test data. We would also like to thank Dan Protopopescu for providing a geometry driver that correctly created the SiD barrel's overlaping module structure, and Marco Oriunno for providing the photo of the prototype calorimeter used in Fig.~\ref{setup}.



\begin{thebibliography}{99}
	\bibitem{tdr4}  T. Behnke, J. Brau \textit{et al.}, ``The Interntional Linear Collider Technical Design Report - Volume 4: Detectors," \href{https://arxiv.org/abs/1306.6329}{\url{arXiv:1306.6329v1 [physics.ins-det]}}
	\bibitem{kpix} J. Brau, M. Breidenbach \textit{et al.}, ``KPiX - A 1,024 Channel Readout ASIC for the ILC," SLAC-PUB-15285 (2013), 2012 IEEE Nuclear Science SYmposium, \url{http://slac.stanford.edu/pubs/slacpubs/15250/slac-pub-15285.pdf}
	\bibitem{lcws13} M. Breidenbach \textit{et al.}, ``Prototype Silicon-tungsten Ecal with Integrated Electronics: First Look with Test Beam," 2013 Linear Collider Workshop (LCWS), Tokyo, \url{https://agenda.linearcollider.org/event/6000/contributions/27576/}
\end{thebibliography}
\end{document}